\documentclass[12pt]{article}

\usepackage{graphicx}

\begin{document}

\title{Geometry of the Vapor Liquid Coexistence in the Gibbs Space}

\author{E Pi\~na\footnote{With sabbatic leave from Universidad Aut\'onoma Metropolitana.} \\{\sl Professor ``Eugenio M\'endez Docurro 2011"} \\de la Escuela Superior de F\'\i sica y Matem\'aticas del IPN, \\Zacatenco 07738, M\'exico D F, Mexico \\
e-mail: pge@xanum.uam.mx}

\date{ }

\maketitle
\center{

}

\abstract{
The phase coexistence is illuminated with geometric views of the thermodynamic variables, according to Gibbs' choices. Quantities and relations between them are obtained. The existence of the edge of regression with tangents coincident with the straight lines connecting the coexistence points of phase equilibrium is stressed. A geometric approach to the critical point leads to estimation of the values of the critical exponents for the angles formed by the coexistence curve and the straight lines with the principal direction along the minimal curvature.}

\newpage

\section{Introduction}
It is astonishing the existence of useful, yet little known properties of the Thermodynamic variables in the case of phase coexistence of a pure substance. In most textbooks the emphasis is on the knowledge of the vapor pressure as function of the temperature, the heat of vaporization, the discontinuity in the molar volume, the Clausius-Clapeyron equation \cite{pr}. All this should be included in the teaching of Classical Thermodynamics. Besides these relevant quantities it is shown here that equally important quantities are the chemical potential at coexistence as a function of the temperature and the discontinuities in the molar energy. The essential point in this paper is to show the use of the Gibbs' geometric space of Entropy, Volume, Energy $(S, V, E)$ to relate the set of the above thermodynamic functions, and it is surprising to find new geometric quantities in terms of known variables that provides new useful properties to be used for teachers to master the science, for students to increase the knowledge of the subject, and to practitioners of the art of Thermodynamics to take advantage of the new properties.

Some of these properties were published originally in Spanish \cite{pm}, and therefore are not well known for many scientists not reading in Spanish language. This is a reason to recover them and extend to new properties.

A remarkable part of the following geometry was described by J. W. Gibbs \cite{gi} and several mathematical proofs were completed by P. Saurel \cite{sa}, nevertheless they were forgotten.

The geometry used in this paper is the ordinary geometry necessary to draw the thermodynamic figures as appear in the thermodynamic texts; as is used in any first course of thermodynamics. This geometry fully agrees with the W. Gibbs \cite{gi} and R. Gilmore \cite{g2} choices of geometry. Do not confuse this elementary geometric approach with those that study a differential geometry with a metric tensor formed with second derivatives of the internal energy; that postulate an analogous treatment of thermodynamics and general relativity \cite{we}. We are very far from those.

Some mathematical care could be included noting that as the different coordinates have not the same physical dimensions, the geometry we use in thermodynamics is actually an affine geometry, as Tisza pointed out \cite{ti}. The mathematical reader understands that this implies we have not some metric properties. Nevertheless the concepts of parallelism and orthogonality are not lost. Also the approach of two points in this space until the two become one, in the limit, is not lost; because the distance along each coordinate axis is not lost. I am assuming the reader is able to understand these geometric concepts as in ordinary geometry, as in a pedestrian way; not with the deep treatment of a mathematic theory. With this perspective one defines many geometric concepts which are elementary concepts for most mathematicians.

Some results in this paper are found in a different thermodynamic variables in \cite{pi}. Notwithstanding this, some proofs presented in this paper have been never published.

The thermodynamic three-dimensional space with cartesian coordinates: molar entropy $S$, molar volume $V$, and molar energy $E$ was introduced by Gibbs \cite{gi} to represent with a point the state of thermodynamic equilibrium of an homogeneous substance like water. The set of equilibrium states forms a surface with coordinates $S$, $V$ and
\begin{equation}
E = U(S, V)\, ,
\end{equation}
where $U(S, V)$ is the (internal) energy as function of entropy and volume. For one phase this is a continuous function, the surface is a smooth surface. The equation of this surface is written as $f(S, V, E) = 0$ if the function $f$ is chosen as
\begin{equation}
0 = f(S, V, E) = U(S, V) - E\, .\label{function}
\end{equation}

Elementary vector geometry shows that the gradient of a function is orthogonal to a related surface in a point. The gradient is computed at the point. The surface is defined equating to zero the function. The coordinates of the point satisfy the equation of the surface. The gradient is the vector formed by the derivatives of the function with respect to the coordinates.

Applying these definitions to our thermodynamic surface it comes that the gradient vector of coordinates
\begin{equation}
 \left( \frac{\partial f}{\partial S}, \frac{\partial f}{\partial V}, \frac{\partial f}{\partial E}\right) = \left( \left(\frac{\partial U}{\partial S}\right)_V, \left(\frac{\partial U}{\partial V}\right)_S, - 1 \right)\label{gradient}
\end{equation}
is orthogonal to the thermodynamic surface.

Thermodynamics show the Gibbs relation between the thermodynamic surface and the temperature $T$ and pressure $P$: $d U = T dS - P dV$ that means
\begin{equation}
T = \left(\frac{\partial U}{\partial S}\right)_V \, , \qquad - P = \left(\frac{\partial U}{\partial V}\right)_S\, .
\end{equation}
Therefore the temperature and the pressure of the state of equilibrium are determined by these two functions of coordinates $S$ and $V$. And hence the components of the vector (\ref{gradient}) orthogonal to the surface have physical meaning
\begin{equation}
 \left( \frac{\partial f}{\partial S}, \frac{\partial f}{\partial V}, \frac{\partial f}{\partial E}\right) = \left( \left(\frac{\partial U}{\partial S}\right)_V, \left(\frac{\partial U}{\partial V}\right)_S, - 1 \right) = \left(T, - P, - 1 \right)\, .\label{normal}
\end{equation}
Nevertheless the concept of orthogonality has been introduced here assuming the similar concept as was learned in an elementary course on vector (algebra and) analysis. A clear definition comes in what follows.

A curve in the thermodynamic space in parametric form is defined by three functions of a parameter $t$: $S_1(t), V_1(t), E_1(t)$. A point of the curve corresponding to a particular value of the parameter $t$ is obtained computing the three functions for the value of the parameter to obtain the coordinates of the point as
\begin{equation}
S = S_1(t) \, , \qquad V = V_1(t)\, , \qquad E = E_1(t)\, .\label{curve}
\end{equation}

The curve is on the thermodynamic surface (\ref{function}) if substitution of the coordinates $S, V, E$ by the functions $S_1(t), V_1(t), E_1(t)$ in (\ref{function}) gives the identity
\begin{equation}
E_1(t) - U(S_1(t), V_1(t)) = 0 \qquad \mbox{(valid for all $t$).}
\end{equation}
In what follows I use both curves on and outside the thermodynamic surface.

A tangent vector to the curve (\ref{curve}) is defined by the components formed by the derivatives of the functions defining the curve namely
\begin{equation}
\left( \frac{d S_1(t)}{d t}, \frac{d V_1(t)}{d t}, \frac{d E_1(t)}{d t}\right)\, .
\end{equation}

If the curve is on the thermodynamic surface, the tangent vector to the curve is also a tangent vector to the surface.
\begin{equation}
\left( \frac{d S_1(t)}{d t}, \frac{d V_1(t)}{d t}, \frac{d E_1(t)}{d t}\right) = \left( \frac{d S_1(t)}{d t}, \frac{d V_1(t)}{d t}, T \frac{d S_1(t)}{d t} - P \frac{d V_1(t)}{d t}\right)\, .\label{tangent}
\end{equation}

The two vectors (\ref{normal}) and (\ref{tangent}) are orthogonal. Two vectors $(A, B, C)$ and $(a, b ,c)$ are orthogonal when the internal product of the two vectors $(A, B, C) \cdot (a, b ,c) \equiv A a + B b + C c$ is equal to zero. Vector (\ref{normal}) is called orthogonal to the surface since it is orthogonal to the tangent vector of any curve at a point on the surface.

A basis of two tangent vectors to the surface are defined by partial derivatives of the position vector of the surface with respect to the coordinates
$$
{\bf e}_S = \frac{\partial}{\partial S} \, (S, V, U(S, V) = (1, 0, T) \, ,
$$
\begin{equation}
{\bf e}_V = \frac{\partial}{\partial V} \, (S, V, U(S, V) = (0, 1, - P) \, .\label{basis}
\end{equation}
In terms of this basis, vector (\ref{tangent}) has the components $\frac{d S_1(t)}{d t}$ and $\frac{d V_1(t)}{d t}$ namely
\begin{equation}
\left( \frac{d S_1(t)}{d t}, \frac{d V_1(t)}{d t}, T \frac{d S_1(t)}{d t} - P \frac{d V_1(t)}{d t}\right) = \frac{d S_1(t)}{d t} {\bf e}_S + \frac{d V_1(t)}{d t} {\bf e}_V \, .\label{11}
\end{equation}

This basis determines the first fundamental form of the surface $\mathcal{A}$ as follows
\begin{equation}
\mathcal{A} = \left( \begin{array}{cc}
{\bf e}_S \cdot {\bf e}_S & {\bf e}_S \cdot {\bf e}_V \\
{\bf e}_V \cdot {\bf e}_S \cdot  & {\bf e}_V \cdot {\bf e}_V \\
\end{array}\right) = \left( \begin{array}{cc}
1 + T^2 & - T P \\
- T P & 1 + P^2
\end{array}\right) \, .
\end{equation}

Another geometric concept to be used later in this paper is the parallelism of two vectors $(A, B, C)$ and $(a, b ,c)$; when the corresponding components are proportional, namely $A = k a$, and $B = k b$, and $C = k c$ for a given non zero number $k$.

One uses the unit normal vector to the surface parallel to the gradient (\ref{normal})
\begin{equation}
{\bf g} = \frac{1}{\sqrt{1 + T^2 + P^2}} (T, - P, - 1) \, , \qquad {\bf g} \cdot {\bf g} = 1 \, .
\end{equation}

The second fundamental form of the surface $\mathcal{B}$ is defined by
\begin{equation}
\mathcal{B} = \left(  \begin{array}{cc}
{\bf g} \cdot \frac{\partial {\bf e}_S}{\partial S} & {\bf g} \cdot \frac{\partial {\bf e}_S}{\partial V} \\
{\bf g} \cdot \frac{\partial {\bf e}_V}{\partial S} & {\bf g} \cdot \frac{\partial {\bf e}_V}{\partial V}
\end{array}\right) =
\frac{1}{\sqrt{1 + T^2 + P^2}} \left(  \begin{array}{cc}
\left( \frac{\partial T}{\partial S}\right)_V & \left( \frac{\partial T}{\partial V}\right)_S \\
- \left( \frac{\partial P}{\partial S}\right)_V & - \left( \frac{\partial P}{\partial V}\right)_S
\end{array}\right) \, ,
\end{equation}
which is a symmetric matrix as a consequence of the Maxwell relation of thermodynamics
$$
\left( \frac{\partial T}{\partial V}\right)_S = - \left( \frac{\partial P}{\partial S}\right)_V \, .
$$
The introduction of the first and second forms $\mathcal{A}$ and $\mathcal{B}$ of differential geometry has the purpose of introducing the definition of principal curvatures, which must be positive in order to comply with the thermodynamic criteria of stability. Furthermore they are fundamental quantities to study the behavior of the thermodynamic surface near the critical point.

One needs a tool discovered by Euler \cite{eu} in 1760. He finds vectors ${\bf d}_i$  in the basis (\ref{basis}) and curvatures $\lambda_i, (i = 1, 2)$, which are respectively eigenvectors and eigenvalues of the eigenvalue equation
\begin{equation}
\mathcal{B}\, {\bf d}_i = \lambda_i \, \mathcal{A} \ {\bf d}_i \qquad (i = 1, 2)\, ,\label{euler0}
\end{equation}
where $\lambda_i$ are the solutions to the characteristic equation formed with the determinant
\begin{equation}
| \mathcal{B} - \lambda \mathcal{A} | = 0\, .
\end{equation}
The two solutions $\lambda = \lambda_1, \lambda_2$ are the principal curvatures on each point of the surface. In general the two are positive in Thermodynamics. With no loss of generality the principal directions are assumed to be normalized; and Euler proved the two directions to be orthogonal with respect to matrix $\mathcal{A}$ namely
\begin{equation}
{\bf d}^{\rm T}_i \mathcal{A} \, {\bf d}_i = 1\, , \qquad {\bf d}^{\rm T}_1 \mathcal{A} \, {\bf d}_2 = 0\, .\label{euler1}
\end{equation}

Any tangent unit vector $\bf u$ in the surface is written in terms of the angle $\phi$ it forms with the principal direction ${\bf d}_1$ as
\begin{equation}
{\bf u} = {\bf d}_1 \cos \phi + {\bf d}_2 \sin \phi\, .
\end{equation}

Using (\ref{euler0}) and (\ref{euler1}) one has the Euler equation \cite{gu} for the curvature along $\bf u$
\begin{equation}
{\bf u}^{\rm T} \, \mathcal{B} \, {\bf u} = \lambda_1 \cos^2 \phi + \lambda_2 \sin^2 \phi\, .\label{euler2}
\end{equation}
Which means: if $\lambda_1 < \lambda_2$, the curvature is minimum, $\lambda_1$, for $\phi = 0$, and maximum, $\lambda_2$, for $\phi = \pi/2$.

Useful quantities are the Gaussian curvature $K$ equal to the product of the two principal curvatures and written in terms of thermodynamic quantities as
\begin{equation}
K = \lambda_1 \lambda_2 = | \mathcal{B} | / | \mathcal{A} | = \frac{1}{1 + T^2 + P^2} \frac{- \left( \frac{\partial P}{\partial V} \right)_T}{\quad \left( \frac{\partial S}{\partial T} \right)_V} \label{gauss}
\end{equation}
and the mean curvature $H$
$$
2 H =  \lambda_1 + \lambda_2 = {\rm Trace}(\mathcal{A}^{-1} \, \mathcal{B}) =
$$
\begin{equation}
\frac{1}{(1+T^2 + P^2)^{3/2}} \left[(1+P^2) \left(\frac{\partial T}{\partial S}\right)_V+2 T P \left( \frac{\partial T}{\partial V}\right)_S-(1+T^2) \left( \frac{\partial P}{\partial V}\right)_S \right] \, .\label{mean}
\end{equation}

The chemical potential is a thermodynamic quantity defined by the expression
\begin{equation}
\mu = U + P V - T S\, .
\end{equation}
It is expressed in terms of the scalar product of the normal vector (\ref{normal}) and the position vector $(S, V, U)$ of a point on the surface as
\begin{equation}
\mu = - (T, - P, -1) \cdot (S, V, U)\, .\label{potential}
\end{equation}
In this section $U, P, T$ are assumed functions of $S$ and $V$. Therefore $\mu$ is also a function of these variables $S$ and $V$.

\section{Coexisting phases}
A liquid and his vapor are found in thermodynamic equilibrium with characteristic properties. The equilibrium is represented in the Gibbs' space by two corresponding points showing a difference in the values of entropy, volume and energy. However the vapor-liquid equilibrium has in both phases the same temperature, pressure and chemical potential. We have nine thermodynamic functions: the entropy, volume and energy of the liquid; the entropy, volume and energy of the vapor; the temperature, the pressure and the chemical potential. All nine quantities are continuous functions of only one variable. In this section the independent variable is chosen to be the temperature. The entropy, volume, and energy of the vapor as a function of temperature forms a curve on the thermodynamic surface with position vector
\begin{equation}
(S_G(T), V_G(T), U_G(T))\, \qquad \mbox{with} \qquad U_G(t) \equiv U(S_G(T), V_G(T))\, .\label{vapor}
\end{equation}
The entropy, volume, and energy of the liquid as a function of temperature forms a curve on the thermodynamic surface with position vector
\begin{equation}
(S_L(T), V_L(T), U_L(T))\, \qquad \mbox{with} \qquad U_L(t) \equiv U(S_L(T), V_L(T))\, .\label{liquid}
\end{equation}
To that value of temperature correspond a value of the vapor pressure $P(T)$ and of the vapor chemical potential $\mu(T)$. The two curves (\ref{vapor}) and (\ref{liquid}) are boundaries of the thermodynamic surface, which is broken between the two curves forming a gap. On the opposite side to the gap the vapor curve is connected to points of the thermodynamic surface with only the gas phase, which has two independent coordinates $S$ and $V$. On the opposite side to the gap, the liquid curve is connected to points of the thermodynamic surface with only the liquid phase, with its two independent coordinates $S$ and $V$. The normal vector to the surface (\ref{normal}) is defined for the corresponding points of the curves (\ref{vapor}) and (\ref{liquid}) on the surface. This vector is only a function of the temperature and has the same value for the corresponding points in both curves. In terms of the defined notation this vector becomes
\begin{equation}
(T, - P(T), - 1)\, .
\end{equation}

The chemical potential has the same value in the two coexisting phases and is a function of the temperature. The definition (\ref{potential}) for the chemical potential has two expressions namely
\begin{equation}
\mu(T) = U_G(T) + P(T) V_G(T) - T S_G(T) = U_L(T) + P(T) V_L(T) - T S_L(T)\, .\label{vapor-mu}
\end{equation}
The difference of the two last expressions is expressed as the orthogonality of the normal vector and the vector joining the two corresponding points at both sides of the gap
\begin{equation}
0 = \left(T, - P(T) - 1\right) \cdot \left(S_G(T) - S_L(T), V_G(T) - V_L(T), U_G(T) - U_L(T)\right)\, .\label{orthogonal}
\end{equation}

The next geometric idea is to introduce the plane in the Gibbs' space tangent to the thermodynamic surface at the liquid coexisting point of temperature $T$.
The equation of the plane is obtained by substitution of coordinates $S_G(T), V_G(T), U_G(T)$ by a point $S, V, E$ in the previous equation (\ref{orthogonal}).
\begin{equation}
 0 = \left(T, - P(T) - 1\right) \cdot \left(S_L(T) - S, V_L(T) - V, U_L(T) - E\right)\, .\label{plane0}
\end{equation}
The vector from the point $S_L(T), V_L(T), U_L(T)$ to any point $S, V, E$ in the plane is orthogonal to the vector normal to the surface at the same point.
However using one of the expressions for the vapor chemical potential (\ref{vapor-mu}) the equation for this plane is transformed into
\begin{equation}
 0 = - T S + P(T) V + E - \mu(T)\, .\label{plane1}
\end{equation}
For a given $T$ this is the equation of a plane in the Gibbs' space. Changing $T$ it comes a one-parameter family of planes.
This proof is independent of the label for the liquid or the vapor points of coexistence. The plane (\ref{plane1}) is therefore tangent to the two points, it contains the line joining this points at coexistence. This proof is essentially the same that in \cite{sa}, inspired in Gibbs \cite{gi}.

At this point it is interesting to copy a text from Gibbs \cite{gi} who gives the clear geometric picture of this coexistence of phases
\begin{quotation}
...as the tangent plane rolls upon the primitive surface, the two points of contact approach one another and finally fall together. The rolling of the double tangent plane necessarily comes to an end. The point where the two points of contact fall together is the {\sl critical point}...
\end{quotation}

\begin{figure}

\centering

\scalebox{0.45}{\includegraphics[angle=90]{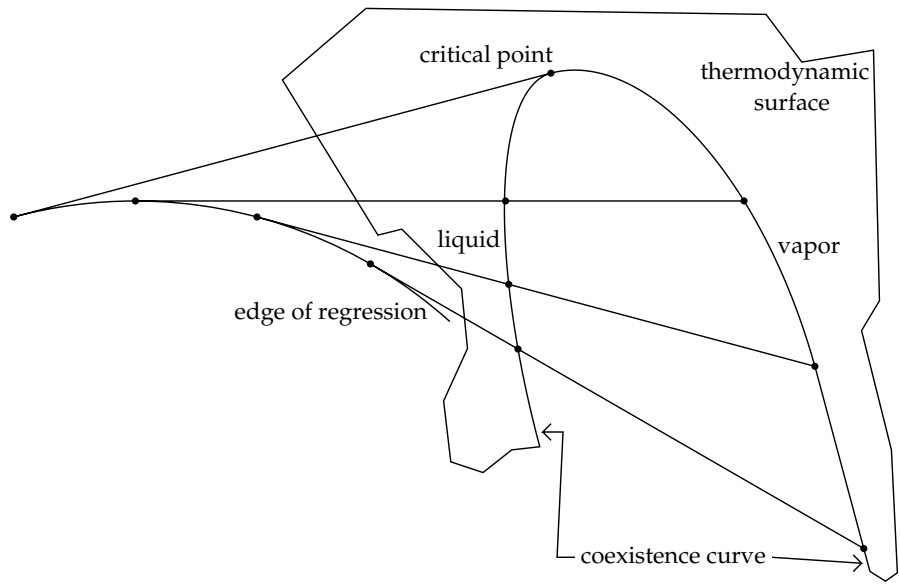}}

\caption{Curves liquid and vapor of coexistence of phases. The coexisting points are joined by a straight line tangent to the edge of regression.}
\end{figure}

The envelope of the family of planes that rolls on the thermodynamic surface is a developable surface formed by the straight lines joining the coexisting states of phase equilibrium. The straight lines are at the intersection of the tangent plane (\ref{plane1}) and a neighboring plane at $T + dT$. These two planes intersect on the straight line and with the plane also, obtained from (\ref{plane1}) by taking a partial derivative with respect to the parameter $T$, namely
\begin{equation}
0 = - S + P'(T) V  - \mu'(T)\, .\label{plane2}
\end{equation}
That the coexistence points are on this plane comes from the Gibbs-Duhem relation on both points of the coexistence curve
\begin{equation}
\mu' = - S_A + V_A P' \qquad (A = L, G) \, .\label{gd}
\end{equation}
In what follows equations with subindex $A$ represent two equations, one for $A=L$ and other for $A=G$.

The cross product of the directed vectors of the planes (\ref{plane1}-\ref{plane2})
$$
\left(T, - P(T), - 1\right) \qquad \mbox{and} \qquad \left(1, - P'(T), 0 \right)
$$
is the vector
\begin{equation}
 \left( P'(T), 1, - P(T) + T P'(T) \right) \, ,\label{tan}
\end{equation}
which should be parallel to the vector joining the two corresponding points at both sides of the gap
\begin{equation}
 \left(S_G(T) - S_L(T), V_G(T) - V_L(T), U_G(T) - U_L(T)\right)\, ,
\end{equation}
therefore the important result
$$
\left(S_G(T) - S_L(T), V_G(T) - V_L(T), U_G(T) - U_L(T)\right) =
$$
\begin{equation}
 [V_G(T) - V_L(T)] \left( P'(T), 1, - P(T) + T P'(T) \right)\, ,
\end{equation}
comes about implying the Clausius Clapeyron equation
\begin{equation}
P'(T) = \frac{S_G(T) - S_L(T)}{V_G(T) - V_L(T)}
\end{equation}
and
\begin{equation}
 - P(T) + T P'(T) = \frac{U_G(T) - U_L(T)}{V_G(T) - V_L(T)} \, .
\end{equation}

Classical books on differential geometry \cite{ei} teach us about three types of developable surfaces: cylinders, cones and surfaces tangent to a space curve. Experiment allows to discard the cone and cylinder cases, and one focus on the curve called {\sl edge of regression} whose straight line tangents are the straight lines of the developable surface connecting the coexistence points. The edge of regression is the envelope of the straight lines at the intersection of the planes (\ref{plane1}) and (\ref{plane2}). Two neighbor straight lines cut at $T$ and $T + dT$. The position of the edge of regression is then at the intersection of these two planes and a third plane obtained from the partial derivative of the equation of the plane (\ref{plane2}) with respect to $T$ namely
\begin{equation}
0 = P''(T) V  - \mu''(T)\, .\label{plane3}
\end{equation}
From the equations of these three planes one has the coordinates $(S, V, E)$ of the edge of regression
\begin{equation}
\left(\frac{\mu''(T)}{P''(T)} P'(T) - \mu'(T) , \frac{\mu''(T)}{P''(T)}, [T P'(T) - P(T)] \frac{\mu''(T)}{P''(T)} - T \mu'(T) + \mu(T) \right)\, .
\end{equation}

Notice that this point is aligned with the two coexisting points. A tangent vector to the edge of regression is the derivative of this vector with respect to $T$ which is parallel to the vector (\ref{tan}).

The position of any branch of the coexisting curve may be expressed as the position of the edge of regression plus a multiple of the tangent vector (\ref{tan}) to the edge of regression. Some simplification result if we use the volume in the following expression
$$
\left( S_A(T), V_A(T), U_A(T) \right) = \left( - \mu'(T) , 0, - T \mu'(T) + \mu(T) \right) +
$$
\begin{equation}
 V_A(T) \left( P'(T), 1, - P(T) + T P'(T) \right)\, .
\end{equation}

The derivative with respect to the temperature of this coexistence curve is a tangent vector to such curve which is expressed in the form valid for both branches liquid and vapor
$$
\left( S'_A(T), V'_A(T), U'_A(T) \right) = [- \mu''(T) + V_A(T) P''(T)]\left( 1 , 0, T \right) +
$$
\begin{equation}
 V'_A(T) \left( P'(T), 1, - P(T) + T P'(T) \right)\, .
\end{equation}

On approaching the critical point all terms in this equation become divergent. To manage finite quantities divide both members by $V'_A$ to find a finite vector tangent to the curves of coexistence
\begin{equation}
 \left( \frac{S'_A}{V'_A}, 1, \frac{U'_A}{V'_A} \right) = \frac{- \mu'' + V_A P''}{V'_A} \left( 1 , 0, T \right) + \left( P', 1, - P + T P' \right)\, , \label{identity}
\end{equation}
where the explicit dependence of all the quantities as functions of the temperature is no longer explicit. This equation is a thermodynamic identity with a non trivial meaning that is discovered as follows. Deriving the Gibbs-Duhem relation (\ref{gd}) with respect to the temperature, it becomes
$$
\mu'' = - S_A' + V_A' P' + V_A P''
$$
which is rewritten by using the chain rule as
$$
\mu'' -  V_A P''= - S_A' + V_A' P' =
$$
\begin{equation}
- \left[ \left( \frac{\partial S}{\partial T}\right)_V + \left( \frac{\partial S}{\partial V}\right)_T V_A'\right] + V_A' \left[ \left( \frac{\partial P}{\partial T}\right)_V + \left( \frac{\partial P}{\partial V}\right)_T V_A'\right] \, .
\end{equation}
A Maxwell relation allow to simplify it to
\begin{equation}
\mu'' -  V_A P''= - S_A' + V_A' P' = - \left( \frac{\partial S}{\partial T}\right)_V + \left( \frac{\partial P}{\partial V}\right)_T (V_A')^2 \, .\label{44}
\end{equation}
This quantity. changed of sign and divided by $V_A'$ gives a quantity on the right hand of equation (\ref{identity})
\begin{equation}
\frac{- \mu'' +  V_A P''}{V_A'}=  \frac{S_A'}{V_A'} - P' = \frac{1}{V_A'} \left( \frac{\partial S}{\partial T}\right)_V - \left( \frac{\partial P}{\partial V}\right)_T V_A' \, .
\end{equation}
Then equation (\ref{identity}) becomes
\begin{equation}
\left( \frac{S'_A}{V'_A}, 1, \frac{U'_A}{V'_A} \right) = \left[ \frac{1}{V_A'} \left( \frac{\partial S}{\partial T}\right)_V - \left( \frac{\partial P}{\partial V}\right)_T V_A'\right] \left( 1 , 0, T \right) + \left( P', 1, - P + T P' \right)\, .\label{pina}
\end{equation}

Direction $\left( S'_A(T), V'_A(T), U'_A(T) \right)$ is a particular case of a curve tangent to the thermodynamic surface satisfying equation (\ref{11})
\begin{equation}
\left( S'_A(T), V'_A(T), U'_A(T) \right) = S'_A {\bf e}_S + V'_A {\bf e}_V \, .
\end{equation}
which is a tangent vector to the coexistence curve $A = L, G$.

Direction$\left( P'(T), 1, - P(T) + T P'(T) \right)$ in the basis (\ref{basis}) is
\begin{equation}
\left( P'(T), 1, - P(T) + T P'(T) \right) = P'(T) {\bf e}_S + {\bf e}_V \, .
\end{equation}
which is a tangent vector tangent to the edge of regression and tangent to the thermodynamic surface at the two points of coexistence.
These two directions are conjugated at any point of the curve of coexistence. Eisenhart \cite{ei} call them conjugated directions when they are orthogonal with respect to the second fundamental form of the surface, namely
\begin{equation}
\left( P' \ 1\right) \mathcal{B} \left(\begin{array}{c}
S'_A \\
V'_A
\end{array} \right) = 0\, .\label{conjug}
\end{equation}
To prove this equation we multiply the matrix in the definition of $\mathcal{B}$ by the vector in the tangent direction to the coexistence curve and it becomes
\begin{equation}
\left(  \begin{array}{cc}
\left( \frac{\partial T}{\partial S}\right)_V & \left( \frac{\partial T}{\partial V}\right)_S \\
- \left( \frac{\partial P}{\partial S}\right)_V & - \left( \frac{\partial P}{\partial V}\right)_S
\end{array}\right) \left(\begin{array}{c}
S'_A \\
V'_A
\end{array} \right) = \left( \begin{array}{c}
1 \\
-P'
\end{array}
\right)\, ,\label{fiA}
\end{equation}
where we used the chain rule. Equation (\ref{conjug}) is now obviously satisfied.

In order to use the Euler equation (\ref{euler2}) for these two directions let introduce ${\bf u}_R$ the unit vector along the tangent to the edge of regression; and ${\bf u}_A$ the unit vector along the direction of the tangent to the curve of coexistence ($A = L \mbox{ or } G$). The angles of these directions with respect to the principal direction ${\bf d}_1$ are respectively $\phi_R$ and $\phi_A$
\begin{equation}
{\bf u}_R = {\bf d}_1 \cos \phi_R + {\bf d}_2 \sin \phi_R\, , \qquad {\bf u}_A = {\bf d}_1 \cos \phi_A + {\bf d}_2 \sin \phi_A\, .\label{angles}
\end{equation}
The corresponding Euler equations are
\begin{equation}
{\bf u}_R^{\rm T} \, \mathcal{B} \, {\bf u}_R = \lambda_1 \cos^2 \phi_R + \lambda_2 \sin^2 \phi_R\, ,\label{euler4}
\end{equation}
and
\begin{equation}
{\bf u}_A^{\rm T} \, \mathcal{B} \, {\bf u}_A = \lambda_1 \cos^2 \phi_A + \lambda_2 \sin^2 \phi_A\, ,\label{euler5}
\end{equation}

Substitution of equations (\ref{angles}) in the orthogonality condition of conjugated directions give the property \cite{ei}
\begin{equation}
\tan \phi_R \tan \phi_A = - \frac{\lambda_1}{\lambda_2}\, .\label{conju2}
\end{equation}
Since the principal curvatures are positive the two angles have different sign. The principal direction ${\bf d}_1$ is between the two conjugated directions.

\section{Approaching the critical point}
As the critical point is approached some thermodynamic quantities (some partial derivatives in the two phase region, and some ordinary derivatives in the coexisting region) diverge with different singularities and other remain finite. The partial derivatives with a known behaviour are represented as functions of the relative difference of temperature with respect to the critical temperature $T_C$
\begin{equation}
\tau = \left|\frac{T - T_C}{T_C}\right|
\end{equation}

\noindent Hypothesis 1. The derivative of the entropy with respect to temperature at constant pressure diverges with the exponent $- \gamma$
\begin{equation}
\left( \frac{\partial S}{\partial T} \right)_P \approx \tau^{- \gamma} \, .
\end{equation}

\noindent Hypothesis 2. The derivative of the entropy with respect to temperature at constant volume diverges with the exponent $- \alpha$
\begin{equation}
\left( \frac{\partial S}{\partial T} \right)_V \approx \tau^{- \alpha} \, .
\end{equation}

The $\gamma$ and $\alpha$ are positive numbers. The theoretical and experimental values of these exponents that one finds in the literature satisfy the inequalities
\begin{equation}
\gamma - \alpha \geq 1 \, , \qquad 0 \leq \alpha < 0.2\, .
\end{equation}

\noindent Hypothesis 3. The partial derivative of the pressure with respect to the temperature at constant volume is a finite quantity
\begin{equation}
\left( \frac{\partial P}{\partial T} \right)_V \approx \tau^0 \, .
\end{equation}

\noindent Proposition 1. Using thermodynamic identities one finds the behaviour of the other partial derivatives between the four variables $S, V, T, P$.

From Maxwell relation
\begin{equation}
\left( \frac{\partial P}{\partial T} \right)_V = \left( \frac{\partial S}{\partial V} \right)_T \qquad \rightarrow \qquad \left( \frac{\partial S}{\partial V} \right)_T \approx \tau^0 \, .
\end{equation}
From the identity
\begin{equation}
\left( \frac{\partial S}{\partial T} \right)_V = \left( \frac{\partial S}{\partial P} \right)_V \left( \frac{\partial P}{\partial T} \right)_V \qquad \rightarrow \qquad \left( \frac{\partial S}{\partial P} \right)_V \approx \tau^{- \alpha} \, .
\end{equation}
From Maxwell relation
\begin{equation}
\left( \frac{\partial S}{\partial P} \right)_V = - \left( \frac{\partial V}{\partial T} \right)_S \qquad \rightarrow \qquad \left( \frac{\partial V}{\partial T} \right)_S \approx \tau^{- \alpha} \, .
\end{equation}
From chain rule and a Maxwell relation
\begin{equation}
\left( \frac{\partial S}{\partial T} \right)_P - \left( \frac{\partial S}{\partial T} \right)_V = \left( \frac{\partial P}{\partial T} \right)_V \left( \frac{\partial V}{\partial T} \right)_P \qquad \rightarrow \qquad \left( \frac{\partial V}{\partial T} \right)_P \approx \tau^{- \gamma} \, . \label{chain}
\end{equation}
From Maxwell relation
\begin{equation}
\left( \frac{\partial V}{\partial T} \right)_P = - \left( \frac{\partial S}{\partial P} \right)_T \qquad \rightarrow \qquad \left( \frac{\partial S}{\partial P} \right)_T \approx \tau^{- \gamma} \, .
\end{equation}
From the identity
\begin{equation}
\left( \frac{\partial S}{\partial T} \right)_P = \left( \frac{\partial S}{\partial V} \right)_P \left( \frac{\partial V}{\partial T} \right)_P \qquad \rightarrow \qquad \left( \frac{\partial S}{\partial V} \right)_P \approx \tau^0 \, .
\end{equation}
From Maxwell relation
\begin{equation}
\left( \frac{\partial S}{\partial V} \right)_P = - \left( \frac{\partial P}{\partial T} \right)_S \qquad \rightarrow \qquad \left( \frac{\partial P}{\partial T} \right)_S \approx \tau^0 \, .
\end{equation}
From the identity
\begin{equation}
\left( \frac{\partial V}{\partial P} \right)_T = - \left( \frac{\partial V}{\partial T} \right)_P \left( \frac{\partial T}{\partial P} \right)_V \qquad \rightarrow \qquad \left( \frac{\partial V}{\partial P} \right)_T \approx \tau^{- \gamma} \, .
\end{equation}
From the identity
\begin{equation}
\left( \frac{\partial V}{\partial P} \right)_S = \left( \frac{\partial V}{\partial T} \right)_S \left( \frac{\partial T}{\partial P} \right)_S \qquad \rightarrow \qquad \left( \frac{\partial V}{\partial P} \right)_S \approx \tau^{- \alpha} \, .
\end{equation}

\noindent Proposition 2. When we use the chain rule in which two terms have the same exponent $\sigma$, the other term has a larger exponent $\sigma + \gamma - \alpha$.

For example equation (\ref{chain}). Other example is
\begin{equation}
\left( \frac{\partial P}{\partial T} \right)_V = \left( \frac{\partial P}{\partial T} \right)_S + \left( \frac{\partial P}{\partial S} \right)_T \left( \frac{\partial S}{\partial T} \right)_V \, .
\end{equation}
The two terms $\left( \frac{\partial P}{\partial T} \right)_V$ and $\left( \frac{\partial P}{\partial T} \right)_S \approx \tau^0$. The third term $\left( \frac{\partial P}{\partial S} \right)_T \left( \frac{\partial S}{\partial T} \right)_V \approx \tau^{0 + \gamma - \alpha}$\, . Other example is
\begin{equation}
\left( \frac{\partial V}{\partial S} \right)_T = \left( \frac{\partial V}{\partial S} \right)_P + \left( \frac{\partial V}{\partial P} \right)_S \left( \frac{\partial P}{\partial S} \right)_T \, .
\end{equation}
The two terms $\left( \frac{\partial V}{\partial S} \right)_T$ and $\left( \frac{\partial V}{\partial S} \right)_P \approx \tau^0$. The third term $\left( \frac{\partial V}{\partial P} \right)_S \left( \frac{\partial P}{\partial S} \right)_T \approx \tau^{0 + \gamma - \alpha}$\, . Other example is
\begin{equation}
\left( \frac{\partial V}{\partial T} \right)_P = \left( \frac{\partial V}{\partial T} \right)_S + \left( \frac{\partial V}{\partial S} \right)_T \left( \frac{\partial S}{\partial T} \right)_P \, .
\end{equation}
The two terms $\left( \frac{\partial V}{\partial T} \right)_P$ and $\left( \frac{\partial V}{\partial S} \right)_T \left( \frac{\partial S}{\partial T} \right)_P \approx \tau^{- \gamma}$. The third term $\left( \frac{\partial V}{\partial T} \right)_S  \approx \tau^{- \gamma + \gamma - \alpha}$\, .

\noindent Hypothesis 4. On the coexistence curve the derivative of the volume with respect to the temperature diverge as
\begin{equation}
V'_A \approx \tau^{\beta - 1} \, , \qquad (\beta \doteq 1/3) \, .
\end{equation}

\noindent Proposition 3. The derivative of the pressure vapor with respect to the temperature is a finite quantity near the critical point. Using the chain rule one has
\begin{equation}
P' = \left( \frac{\partial P}{\partial T} \right)_V + \left( \frac{\partial P}{\partial V} \right)_T V'_A \, .
\end{equation}
The first term of the right hand side is finite: $\left( \frac{\partial P}{\partial T} \right)_V \approx \tau^0$. The second term goes to cero: $\left( \frac{\partial P}{\partial V} \right)_T V'_A  \approx \tau^{\gamma + \beta - 1}$. Theory and experiment agree with the value $\gamma + \beta - 1 \geq 1/2$.

\noindent Proposition 4. On the coexistence curve the derivative of the entropy with respect to temperature diverge as $\tau^{\beta - 1}$.  Using the chain rule one has
\begin{equation}
S'_A = \left( \frac{\partial S}{\partial T} \right)_V + \left( \frac{\partial S}{\partial V} \right)_T V'_A \, .
\end{equation}
The first term of the right hand side diverge $\left( \frac{\partial S}{\partial T} \right)_V \approx \tau^{- \alpha}$, the second term $\left( \frac{\partial S}{\partial V} \right)_T V'_A \approx \tau^{\beta - 1}$. This last is the dominant term since theory and experiment agree that $1 - \alpha - \beta \geq 1/2$.

\noindent Proposition 5. Among these critical exponents, Rushbrooke \cite{ru} introduces the inequality
\begin{equation}
\alpha + 2 \beta + \gamma \geq 2 \, .
\end{equation}
This is also written in redundant form as
\begin{equation}
\gamma + \beta -1 \geq \frac{\gamma - \alpha}{2} \geq 1 - \alpha - \beta \, .
\end{equation}
Note that these three exponents, separated by inequalities, appear when using the chain rule. The middle exponent is the average of the other two. The three are assumed equal in many cases, and the last is always $\geq 1/2$. We will use the notation
\begin{equation}
\Delta = \alpha + 2 \beta + \gamma -2 \geq 0\, .
\end{equation}

Taking the expression for the (double of) the mean curvature $\lambda_1 + \lambda_2$ in (\ref{mean}) it results that it goes to zero with exponent $\alpha$. The Gaussian curvature, according to (\ref{gauss}) goes to zero with exponent $\alpha + \gamma$. Therefore the minimum curvature $\lambda_1  \approx \tau^\gamma$ and the maximum curvature $\lambda_2 \approx \tau^\alpha$.

In equation (\ref{pina}) the term between square brackets is formed by two terms with same sign. Both terms go to zero. The term $\frac{1}{V_A'} \left( \frac{\partial S}{\partial T}\right)_V \approx \tau^{1 - \alpha - \beta}$. The second term $- \left( \frac{\partial P}{\partial V}\right)_T V_A' \approx \tau^{\gamma + \beta -1}$.

Therefore the two conjugated directions become parallel near the critical point. And because the principal direction happens to be between them, the two conjugated directions tend to the ${\bf d}_1$ direction. The two angles $\phi_R$ and $\phi_A$ go to zero. From equation (\ref{conju2}) the tangent functions are replaced by the angles and the right hand side of that equation gives the property
\begin{equation}
\phi_R \phi_A \approx \tau^{\gamma-\alpha}\, .\label{expo1}
\end{equation}

We use equation (\ref{euler5}) to determine the exponent associated to angle $\phi_A$. To compute the exponent associated to the left hand side of this equation, forgetting the finite terms, one uses equation (\ref{fiA})
$$
\left( S'_A/V'_A \ 1 \right) \left(  \begin{array}{cc}
\left( \frac{\partial T}{\partial S}\right)_V & \left( \frac{\partial T}{\partial V}\right)_S \\
- \left( \frac{\partial P}{\partial S}\right)_V & - \left( \frac{\partial P}{\partial V}\right)_S
\end{array}\right) \left(\begin{array}{c}
S'_A/V'_A \\
1
\end{array} \right) =
 $$
\begin{equation}
 \frac{1}{V'_A} (S'_A/V'_A - P') = \frac{1}{V'_A} \left[ \frac{1}{V'_A} \left( \frac{\partial S}{\partial T}\right)_V - \left( \frac{\partial P}{\partial V}\right)_T V_A' \right]\, .
\end{equation}
The two terms within the square brackets were considered before with exponents $1-\alpha-\beta$ and $\gamma+\beta-1$. Divided by $V'_A$ the left hand side of equation (\ref{euler5}) becomes with exponents $\gamma$ and $\gamma - \Delta$. The dominant exponent is $\gamma - \Delta$. To become equal to the right hand side the dominant exponent comes from the term $\lambda_2 \sin^2 \phi_A$ what allows to infer $\phi_A \approx \tau^{(\gamma -\alpha -\Delta)/2}$ and since $(\gamma -\alpha -\Delta)/2 = 1 - \alpha - \beta$
\begin{equation}
\phi_A \approx \tau^{1 - \alpha - \beta}\, .\label{expo2}
\end{equation}

The exponent of angle $\phi_R$ comes from equations (\ref{expo1}) and (\ref{expo2}) as
\begin{equation}
\phi_R \approx \tau^{\gamma + \beta - 1} \, .\label{expo3}
\end{equation}

Verification of this result is obtained from (\ref{euler4}) similar to the previous equation. Now we use the thermodynamic identity
\begin{equation}
\left(  \begin{array}{cc}
\left( \frac{\partial T}{\partial S}\right)_V & \left( \frac{\partial T}{\partial V}\right)_S \\
- \left( \frac{\partial P}{\partial S}\right)_V & - \left( \frac{\partial P}{\partial V}\right)_S
\end{array}\right) =  \frac{- \left( \frac{\partial P}{\partial V} \right)_T}{\quad \left( \frac{\partial S}{\partial T} \right)_V} \left(  \begin{array}{cc}
- \left( \frac{\partial V}{\partial P} \right)_T & \left( \frac{\partial S}{\partial P} \right)_T \\
- \left( \frac{\partial V}{\partial T} \right)_P & \left( \frac{\partial S}{\partial T} \right)_P
\end{array}\right)\, .
\end{equation}
The left hand side of equation (\ref{euler4}), forgetting the finite terms becomes
$$
\frac{- \left( \frac{\partial P}{\partial V} \right)_T}{\quad \left( \frac{\partial S}{\partial T} \right)_V} \left( P' \ 1 \right)\left( \begin{array}{cc}
- \left( \frac{\partial V}{\partial P} \right)_T & \left( \frac{\partial S}{\partial P} \right)_T \\
- \left( \frac{\partial V}{\partial T} \right)_P & \left( \frac{\partial S}{\partial T} \right)_P
\end{array}\right) \left(\begin{array}{c}
P' \\
1
\end{array} \right) = 
$$
\begin{equation}
\frac{- \left( \frac{\partial P}{\partial V} \right)_T}{\quad \left( \frac{\partial S}{\partial T} \right)_V} \left(- V'_A \ S'_A \right) \left(\begin{array}{c}
P' \\
1
\end{array} \right)\, ,
\end{equation}
where we used the chain rule. Similarity with the previous equation and using that the first quotient goes to zero like $K$ i. e. as $\tau^{\gamma+\alpha}$ gives for the left hand side the exponents $\gamma + \Delta$ and $\gamma$. The dominant exponent is $\gamma$. To become equal to the right hand side the dominant exponent comes from the term $\lambda_1 \cos^2 \phi_R$ which is $\gamma$. The term $\lambda_2 \sin^2 \phi_R$, taking in account (\ref{expo3}) give a behavior which goes to zero faster or equal to the dominant term. Nothing new results.

\section*{Acknowledgments}
The Author assisted Professor L. Garc\'\i a-Col\'\i n's lectures on Thermodynamics at Escuela Superior de F\'\i sica y Matem\'aticas (ESFM) in 1962. L. Garc\'\i a-Col\'\i n, now Emeritus Professor of UAM-Iztapalapa, was founder of ESFM at 1961. This paper is inspired by the broad knowledge on critical point and Physics learned from late Professor Melville Green who visited ESFM at those times. Several visits to discuss Physics in Mexico by Professor M. Green, Professor J. V. Sengers and Professor A. Sengers are deeply recognized.

\end{document}